\documentclass[twocolumn,aps,pra]{revtex4}
\usepackage{epsfig}
\usepackage[english]{babel}
\usepackage{latexsym}
\usepackage{graphics}
\usepackage{subfigure}
\usepackage{graphicx}
\usepackage{dcolumn}
\usepackage{amsmath}
\usepackage{hyperref}
\usepackage{amssymb}
\usepackage{color}

\begin{document}
\title{Measuring Coulomb-Induced Ionization Time Lag with a Calibrated Attoclock}

\author{J. Y. Che$^{1,\dag}$, C. Chen$^{1,\dag}$, S. Wang$^{1,2}$, G. G. Xin$^{3}$, and Y. J. Chen$^{1,*}$}

\date{\today}

\begin{abstract}

Electrons in atoms and molecules can not react  immediately to the action of intense laser field.
A time lag (about 100 attoseconds) between instants of the field maximum and the ionization-rate maximum emerges.
This lag characterizes the response time of the electronic wave function to the strong-field ionization event
and has important effects on subsequent ultrafast dynamics of the ionized electron. 
The absolute time lag is not accessible in experiments. 
Here, a calibrated attoclock procedure, which is related to a simple Coulomb-induced temporal correction to  electron trajectories,
is proposed to measure the relative lag of two different ionization events.
Using this procedure, the difference  (i.e., the relative lag) between the ionization time lags of polar molecules in two consecutive half laser cycles
can be probed with a high accuracy.

\end{abstract}

\affiliation{1.College of Physics and Information Technology, Shaan'xi Normal University, Xi'an, China\\
2.School of Physics, Hebei Normal University, Shijiazhuang, China\\
3.School of Physics, Northwest University, Xi'an, China}
\maketitle

\emph{Introduction}.-Whether there is a response time in the interaction between light and atoms or molecules is a basic conceptual problem in quantum mechanics \cite{Muga2,Leone,Maquet,Pazourek}.
Answering this problem is important for understanding  photo-induced physical processes. In particular, it is inevitable in ultrafast science when making efforts to
precisely measure and control the light-assisted electron ultrafast motion. As great research progresses on this issue have been made in recent years \cite{Pazourek},
a definitive answer is still under expectation.
Theoretically, according to the principle of quantum mechanics,  there is no time operator, so the definition of time is difficult in quantum mechanics \cite{Maquet}.
Experimentally, whether the measurement itself will influence the measured process is also a question \cite{Leone}.
However, the definition and measurement of time is unambiguous in classical mechanics \cite{Maquet}.

For the system of an atom or a molecule interacting with a strong infrared (IR) laser pulse, 
the bound electron  can escape from the laser-Coulomb-formed barrier through tunneling \cite{Keldysh,ADK}.
This tunneling event triggers rich physical processes such as above-threshold ionization (ATI) \cite{Agostini1979, Yang1993, Paulus1994, Lewenstein1995,Becker2002} and
high-order harmonic generation (HHG) \cite{McPherson1987, Huillier1991, Corkum,Lewenstein1994}, which have important applications in attosecond science \cite{Krausz2009,Krausz,Vrakking}.
When the laser electric field is strong enough, the behavior of the tunneling electron becomes classical after it exits the barrier \cite{Corkum,Lewenstein1994}.
Therefore, in essence, the strong-field induced ultrafast processes contain the transition
from quantum process (related to tunneling ionization) to classical-like process (related to the electron motion after tunneling).
So, this strong-field system also provides  an ideal platform to investigate the response time. 
Specifically, if there is a response time in the ionization process of this system, 
the response time can be clearly shown by its influence on the subsequent laser-driven classical motion of the escaped electron.

Recent research showed that   \cite{Xie} in the interaction of a strong IR laser pulse  with the He atom,
the time when the laser intensity reaches the peak is inconsistent with the time when the atom has the maximal instantaneous ionization rate.
There is a time lag between the two times. This time lag has a classical correspondence related to
the long-range property of the Coulomb potential which precludes the electron leaving far away from the nuclei instantaneously.
It has a profound influence on the subsequent ultrafast dynamics (such as ATI and HHG) of the electron after tunneling. 
Therefore, it is strongly suggested that there is a response time in strong-laser-matter interaction, 
which is characterized by the Coulomb-induced ionization time lag.

This time lag has shown its significance on attosecond measurement.
According to the electron-trajectory theory \cite{Lewenstein1994,Milosevic2006}
based on strong-field approximation (SFA)   \cite{Faisal,Reiss} where the Coulomb effect is neglected,
both ATI and HHG can be described with complex electron trajectories, which are represented 
by the final (for ATI) or instantaneous (for HHG) momentum of the escaped electron
and the timing of the electron when it leaves or returns to the nuclei. These trajectories build a bridge between the experimental observables  and the desired temporal information of relevant dynamical processes,
and are the theoretical foundation of attosecond measurement.
The Coulomb-modified SFA (MSFA) which considers the effect of long-range Coulomb potential \cite{MishaY,Goreslavski,yantm2010},
however, showed that the Coulomb-induced ionization time lag leads to an important temporal correction to ATI and HHG electron trajectories \cite{Xie}. 

It needs complex theories to deduce this time lag from observables. Particularly, the absolute value of this lag deduced depends on the deducing procedure. 
In   \cite{Xie}, with numerical solution of the time-dependent Schr\"{o}dinger equation (TDSE) and MSFA,
this time lag is evaluated through analyzing the time-dependent continuum populations 
and comparing results of long-range Coulomb potential to short-range ones. But there is uncertainty about how to define the continuum state of a time-dependent system. 
However, it is possible to measure the lag difference between the timing of ionization from distinct electronic states in a single laser pulse \cite{Maquet}.
The strong-field ionization dynamics of  polar molecules with a permanent dipole (PD) differ remarkably in the two half cycles of a laser cycle \cite{Wang2020,Che2021}, 
with providing a platform for identifying this lag difference  between diverse laser-dressed electronic states  (see Fig. 1).

The focus of the paper is whether  this lag difference can be identified with a simple approach and high accuracy without the need of solving TDSE or MSFA.
Such approaches are highly desired in experimental studies, especially for complicated atomic and molecular targets for which TDSE and MSFA simulations are not easy to achieve.
In the following, we will show that a calibrated attosecond-clock  procedure based on the use of photoelectron momentum distribution (PMD) in a strong elliptical laser field provides such possibilities.

\emph{Methods}.-We choose the polar molecule HeH$^+$  as the study objective,
which can be manipulated in present strong-filed experiments  \cite{Wustelt}.
Due to the effect of PD, the HeH$^+$ molecule in strong laser fields stretches rapidly toward larger internuclear distances $R$
and the ionization of the molecule also mainly occurs at larger R \cite{Liwy}.
We therefore study the ionization dynamics of HeH$^+$ in elliptical laser fields at a stretched distance of
$R=2$ a.u.  in the Born-Oppenheimer (BO) approximation.
We assume that the molecular axis of HeH$^+$ is located in the $xy$ plane. The TDSE of the polar molecule in two-dimensional cases is solved with the spectral method \cite{Feit}.
Relevant numerical details  are introduced  in \cite{Wang2017}.
Analytically, we use a MSFA model which also includes the PD effect \cite{Dimitrovski}.
For convenience, we call this model   MSFA-PD to differentiate it from the general MSFA without PD.
The details for these strong-field models  can be found in \cite{Wang2020}.
At $R=2$ a.u., the HeH$^+$ system has the ionization potential of $I_p=1.44$ a.u. and the value of PD  calculated is $D=-0.36$ a.u. \cite{Wang2020}.

The elliptical electric field  $\mathbf{E}(t)$
has the  form of $\mathbf{E}(t)=\vec{\mathbf{e}}_{x}E_{x}(t)+\vec{\mathbf{e}}_{y}E_{y}(t)$,
with $E_{x}(t)=f(t){E_0}/{\sqrt{1+\epsilon^2}}\sin(\omega t)$ and $E_{y}(t)=\epsilon f(t){E_0}/{\sqrt{1+\epsilon^2}}\cos(\omega t)$.
Here, $E_0$ is the laser amplitude corresponding to the peak intensity $I$ and $\epsilon$ is the ellipticity.
$\omega$ is the laser frequency and $f(t)$ is the envelope function.  $\vec{\mathbf{e}}_{x}$($\vec{\mathbf{e}}_{y}$) is the unit vector along the $x(y)$ axis.
Here, the value of $\epsilon=0.87$ is used, implying that the component $E_{x}(t)$ dominates in ionization.
We assume that the molecular axis is oriented parallel to $\vec{\mathbf{e}}_{x}$ and the heavier (lighter) nucleus is located on the right (left) side.
We use trapezoidally shaped laser pulses with a total duration of fifteen cycles, which are linearly turned on and off
for three optical cycles, and then kept at a constant intensity for nine additional cycles.

\begin{figure}[t]
\begin{center}
\rotatebox{0}{\resizebox *{8.5cm}{4.5cm} {\includegraphics {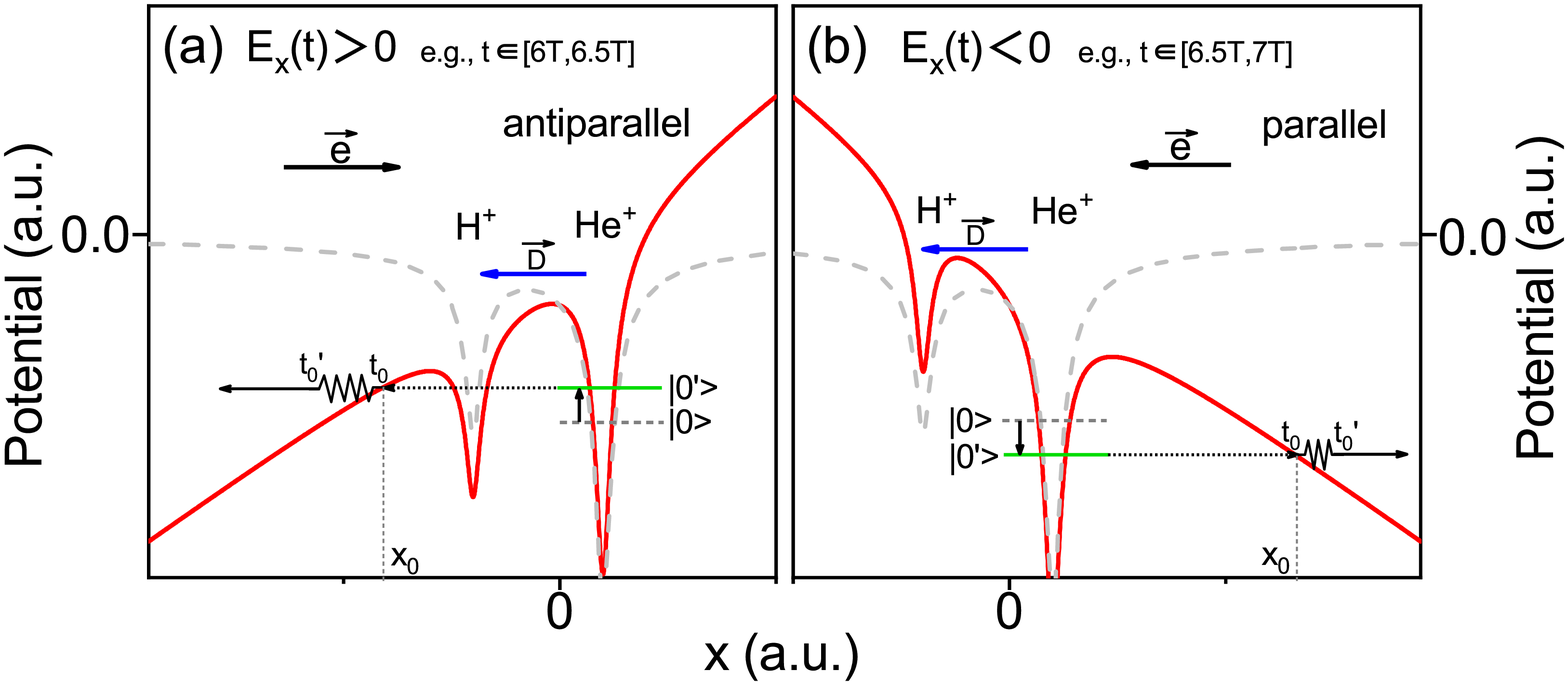}}}
\end{center}
\caption{
Sketch of the Coulomb-induced ionization time lag for the polar molecule HeH$^+$ with a permanent dipole  in one laser cycle.
When the electron exits the laser-Coulomb-formed barrier (red-solid curve) at a time $t_0$, due to the Coulomb effect, it can not be free immediately.
Instead, it stays near the nuclei for a period of $\Delta t$. At the time $t_0'=\Delta t+t_0$,
the instantaneous energy of the electron becomes larger than zero and the electron moves far away from the nuclei. 
For atoms and symmetric molecules, this time lag $\Delta t$  is the same when the tunneling event occurs
in the first (with $E_x(t)>0$) or the second (with $E_x(t)<0$) half cycle of one laser cycle.
For polar molecules, due to the effect of the PD which is directing from He  to H, the situation is different.
When the laser polarization is antiparallel (parallel) to the PD in (a) [(b)],
the energy of the ground state $|0\rangle$ of HeH$^+$ is dressed up (down) and the exit position $x_0$ is nearer to (farther away from) the nuclei.
Accordingly, the Coulomb effect is stronger (weaker) and the time lag $\Delta t$ is larger (smaller) in the first (second) half cycle.
The ionization of HeH$^+$   provides a platform for identifying the difference of $\Delta t$  (i.e., the relative lag) between diverse laser-dressed electronic states.
}
\label{fig:g3}
\end{figure}

\emph{Results and discussions}.-In Fig. 1, we show the dynamics difference
between tunneling ionization of HeH$^+$  in the two half cycles of a laser cycle in the plateau part of the pulse, described with the MSFA-PD.
For the first half laser cycle with $E_x(t)>0$ in Fig. 1(a), the ground state $|0\rangle$ is dressed up and the electron exits the barrier along the H side.
At the exit time $t_0$, the exit position $x_0$ of the tunneling electron is nearer to the nuclei and the electron is subject to a stronger
Coulomb force which precludes the electron to escape. The electron shakes for a while near to the exit position under the action of both the laser field and the Coulomb force,
and up to the time $t_0'$, the instantaneous  energy of the electron including both the kinetic energy and
the potential energy becomes larger than zero, the electron moves far away from the nuclei. Due to the strong Coulomb force at the exit position in this case,
the time difference of $\Delta t=t_0'-t_0$ which is defined as the ionization time lag is also larger. We denote this lag along the H side $\Delta t_{H}$.
The situation reverses for the case of the second half laser cycle in Fig. 1(b), where the Coulomb force is weaker at the exit position and the time lag $\Delta t$ is also smaller.
We denote this lag along the He side $\Delta t_{He}$. When the absolute value of $\Delta t$ depends on the model or the definition used in calculations,
the difference between $\Delta t_{H}$ and $\Delta t_{He}$, which reflects the essential dynamics difference for polar molecules with a large PD in two consecutive half laser cycles,
is expected to be insensitive to those, as to be explored below.

\begin{figure}[t]
\begin{center}
\rotatebox{0}{\resizebox *{8.5cm}{7cm} {\includegraphics {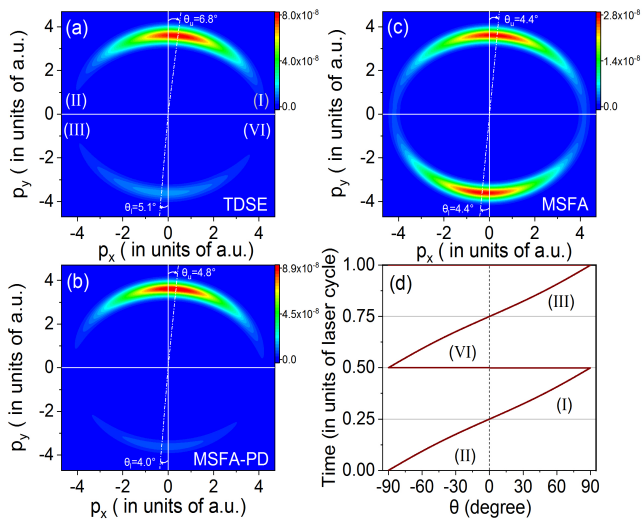}}}
\end{center}
\caption{PMDs of HeH$^{+}$ in elliptical laser fields,
obtained with TDSE (a), MSFA-PD  (b) and  MSFA (c).
The offset angle $\theta$ of PMD arising from the Coulomb effect is defined as shown in (a)-(c).
The offset angle $\theta_u$ in the upper half plane of PMD differs somewhat from the lower one $\theta_l$ in (a) and (b), due to the PD effect. They are the same in (c).
In (d), we plot the curve of the time $t$ as a function of the angle $\theta=\arctan[A_x(t)/A_y(t)]$.
With this curve and considering that the offset angle is only related to the Coulomb-induced ionization time lag, one can deduce this lag  (response time)  directly
from the offset angle of PMD. 
Laser parameters used are $I=2\times10^{15}$W/cm$^{2}$, $\lambda=1000$ nm and $\epsilon=0.87$.
}
\label{fig:g1}
\end{figure}

\begin{figure}[t]
\begin{center}
\rotatebox{0}{\resizebox *{8cm}{7cm} {\includegraphics {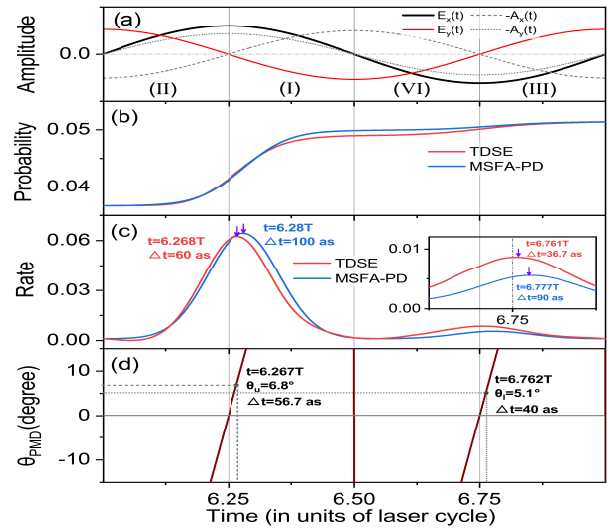}}}
\end{center}
\caption{
The sketch of  electric field $\mathbf{E}(t)$ and  minus vector potential $-\mathbf{A}(t)$ (a), the time-dependent ionization probability (b) and ionization rate (c) of HeH$^+$ calculated with TDSE and
MSFA-PD, and the time-dependent angle $\theta=\arctan[A_x(t)/A_y(t)]$ (d). Results are presented in one laser cycle of $6T$ to $7T$ with $T=2\pi/\omega$.
The one-cycle time region is divided into four parts of I to IV, relating to PMDs in quadrant 1 to quadrant 4.
In (b) and (c),  the model
curves are multiplied by a vertical scaling factor to match the TDSE ones. The inset in (c) shows the enlarged results around $t=6.75T$.
For TDSE and MSFA-PD, the Coulomb-induced ionization time lag $\Delta t$ in the first or the second half laser cycle is defined
as the time difference between the peak times of the ionization rate
and  the electric field $E_x(t)$, as shown in (c).
For CCAC, it is defined as the time difference between the time $t$ with $\arctan(A_x(t)/A_y(t))=\theta_{PMD}$  and the peak time of $E_x(t)$, as shown in (d).
Here, $\theta_{PMD}$ is the offset angle of the upper or the lower half plane of PMD in TDSE simulations.
Laser parameters used are as in Fig. 2.
}
\label{fig:g2}
\end{figure}

\begin{figure}[t]
\begin{center}
\rotatebox{0}{\resizebox *{8.5cm}{7cm} {\includegraphics {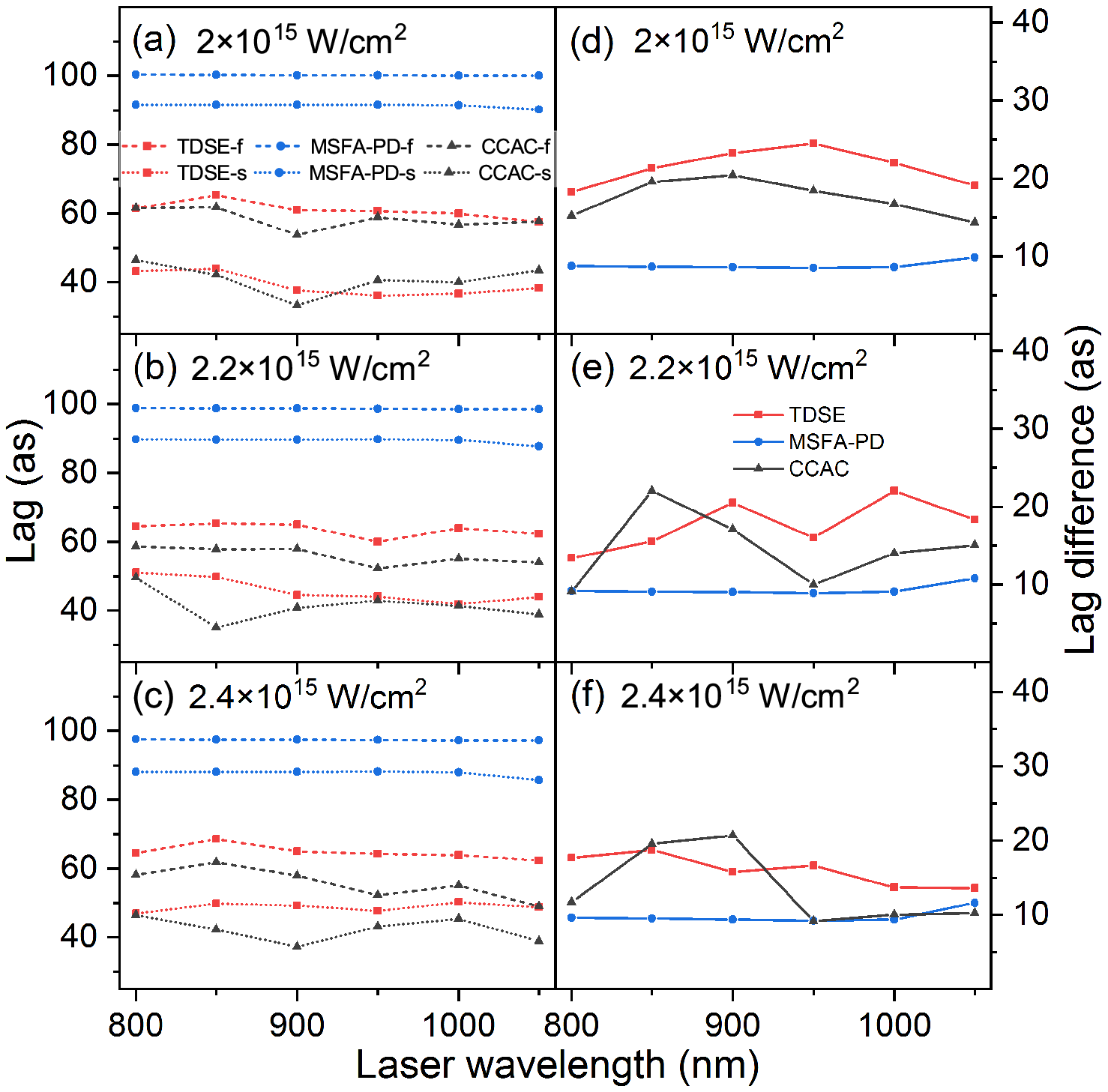}}}
\end{center}
\caption{Time lags of $\Delta t$ (absolute lags, left column) in the first (dashed curves) and the second (dotted curves) half laser cycles for HeH$^+$ and the corresponding lag difference
(relative lag, right column) between these two half laser cycles, calculated with TDSE,
MSFA-PD and CCAC. Laser parameters used are as shown.
}
\label{fig:g4}
\end{figure}

In Fig. 2, we present PMDs of HeH$^{+}$ in elliptical laser fields obtained with different methods.
For shorter laser wavelengthes, some electrons  located in the ground state are pumped into then ionize from the first excited state,
and the excited-state  channel  has an important role in strong-field dynamics of HeH$^+$ \cite{Wang2017}.
Here, we focus on longer wavelengthes (longer than 800 nm) for which the Stark-shifted ground-state channel, as introduced in Fig. 1, dominates in ionization \cite{Che2021}.
For TDSE results in Fig. 2(a), the offset angle $\theta$ ($\theta$,
the angle between the $y$ axis and the axis which goes through the origin and the most bright part of the PMD)  in the upper half plane is
$\theta_u=6.8^o$ and that in the lower half plane is $\theta_l=5.1^o$, with a difference of $\Delta\theta=1.7^o$.
This difference is also reproduced by the MSFA-PD with  $\theta_u=4.8^o$ and $\theta_l=4^o$, as seen in Fig. 2(b).
This difference disappears in the MSFA simulations which consider the asymmetric Coulomb potential of HeH$^+$ but neglect the PD effect, as shown in Fig. 2(c) with $\theta_u=\theta_l=4.4^o$.
The PMDs in the upper and the lower half planes are related to photoelectrons born in the first and the following half laser cycles, respectively (also see Fig. 3(a)).
Without the Coulomb effect, the PMD in elliptical laser fields is symmetric with respect to the $y$ axis and
there is a one-to-one mapping between ionization time and PMD, that is, the PMD acts like an ``attosecond clock", from which one can deduce the ionization-time information with attosecond resolution,
as implied by the classical model \cite{Corkum} or the general SFA \cite{Lewenstein1995}.
The Coulomb effect destroys this symmetry and leads to the nonzero offset angle. As a result, the time information can not be read directly from the PMD.
Semiclassical simulations of strong-field ionization considering the Coulomb effect are therefore needed as a reference to make the time information accessible from the PMD  \cite{Eckle}.
The response time in tunneling ionization, relating to the ionization time lag discussed above, suggests that
the time information can be retrieved from the PMD with considering the response time as a simple temporal correction to the attosecond clock.
In Fig. 2(d), we plot the function curve of the time $t$ versus the angle $\theta$ of the classical prediction defined with $\theta=\arctan[A_x(t)/A_y(t)]$ and $\mathbf{A}(t)=-\int^t\mathbf{E}(t')dt'$.
When the angle is located in the first (I) and the third (III)  (the second (II) and the fourth (IV)) quadrants, we define that it is plus (minus).
Giving the offset angle obtained in TDSE or experiments, one can deduce the time information directly through the function curve.
We call the deduction procedure the Coulomb-calibrated attosecond clock (CCAC).

Next, we perform comparisons between the predictions of TDSE, MSFA and CCAC for this time lag. Relevant results are shown in Fig. 3 and Fig. 4.
The laser electric fields of $E_x(t)$ and $E_y(t)$ and the minus vector potentials of $-A_x(t)=p_x$ and $-A_y(t)=p_y$ in one laser cycle of $6T$ to $7T$ are plotted in the first row of Fig. 3,
with dividing the time region into four parts of I to IV corresponding to PMDs of momenta ($p_x,p_y$) in quadrant 1 to quadrant 4.
The time-dependent ionization probability (continuum population) of TDSE is plotted in the second row of Fig. 2, which shows a remarkable asymmetry
in the first ($6T$ to $6.5T$) versus the second ($6.5T$ to $7T$) half laser cycles. This asymmetry is well reproduced by the MSFA-PD.
The corresponding time-dependent ionization rates (the time derivative of ionization probabilities) are presented in the third row of Fig. 3.
The rate curves of both TDSE and model results  show maxima in each half laser cycle, as indicated by the vertical arrows.
These maxima correspond to the brightest parts of the PMD in the upper and the lower half planes.
Relevant times $t$ related to these maxima are also presented here, which deviate from the instants of $t=6.25T$ and $t=6.75T$  at which the value of $E_x(t)$ reaches its peak.
These times can be understood as the ionization times of photoelectrons with the maximal amplitudes in the first and the following half laser cycles,
and this deviation of these times from the field maxima denotes the Coulomb-induced ionization time lag $\Delta t$.
In the fourth row of Fig. 3, we show the predictions of CCAC for the ionization time $t$ with the offset angles obtained from TDSE.
These ionization times obtained with TDSE, MSFA-PD and CCAC in each half laser cycle differ  from each other,
indicating that the value of the corresponding ionization time lag relative to the peak time of $E_x(t)$ depends on the approach used in evaluations.
However, the situation is different for relative lags, as shown in Fig. 4.

In Fig. 4, we plot the time lags (left column) and the lag differences (right)  in these two half laser cycles for varied laser wavelengthes and intensities, obtained with different methods.
One can observe that when the time lags for different methods and  laser parameters  differ remarkably, the lag differences are located in a time region of 10 to 25 attoseconds.
In particular, the predictions of CCAC for this lag difference are very near to the TDSE ones.
By comparison, the MSFA-PD results deviate from the TDSE predictions with an upper limit of about 10 attoseconds.
This deviation is easily understood, since in the MSFA-PD simulations, the approximate descriptions of the Coulomb and  PD effects both can bring about disagreement with the TDSE.
The similarity between TDSE and CCAC predictions indeed suggests that one can distill the time information from PMDs measured in experiments through CCAC with a high time resolution. 

In summary, we have studied ionization dynamics of HeH$^+$ in strong elliptical IR laser fields.
We have shown that due to the PD effect, the Coulomb induced ionization time lags differ for ionization events occurring
in two consecutive half laser cycles. This lag difference (about 20 attoseconds) is well mapped in the PMD. 
Using a procedure termed as CCAC which considers the Coulomb induced temporal correction to classical predictions, we are able to distill this difference from PMD with a high accuracy. 
The Coulomb induced ionization time lag can be understood as the response time of the electronic wave function to a strong-field ionization event. 
This response time is general and has important effects on tunneling-triggered ultrafast electron dynamics.
Our work provides a feasible manner for evaluating the response time, 
especially for probing the differences of response time between diverse targets and diverse electronic states.

This work was supported by the National Natural Science Foundation of China (Grant No. 91750111),
and the National Key Research and Development Program of China (Grant No. 2018YFB0504400).


\begin{thebibliography}{2}





\bibitem{Muga2} \emph{Time in Quantum Mechanics}, edited by J. G. Muga, R. Sala, and I. L. Egusquiza (Springer, Berlin, 2002).

\bibitem{Leone} S. R. Leone, C. W. McCurdy, J. Burgdorfer, L. S. Cederbaum, Z. Chang, N. Dudovich, J. Feist, C. H. Greene, M. Ivanov, R. Kienberger \emph{et al.},
What will it take to observe processes in `real time'? Nat. Photonics \textbf{8}, 162 (2014).

\bibitem{Maquet} A. Maquet, J. Caillat, and R. Ta\"{\i}eb, Attosecond delays in photoionization: time and quantum mechanics, J. Phys. B \textbf{47}, 204004 (2014).

\bibitem{Pazourek} R. Pazourek, S. Nagele, J. Burgd\"{o}rfer, Attosecond chronoscopy of photoemission, Rev. Mod. Phys. \textbf{87}, 765 (2015).










\bibitem{Keldysh} L. V. Keldysh, Ionization in the field of a strong electromagnetic wave, Sov. Phys. JETP \textbf{20}, 1307 (1965).
\bibitem{ADK} M. V. Ammosov, N. B. Delone, and V. P. Krainov,
Tunnel ionization of complex atoms and of atomic ions in an alternating electric field, Sov. Phys. JETP \textbf{64}, 1191 (1986).




\bibitem{Agostini1979} P. Agostini, F. Fabre, G. Mainfray, G. Petite, and N. K. Rahman, Free-Free Transitions Following Six-Photon Ionization of Xenon Atoms, Phys. Rev. Lett. \textbf{42}, 1127 (1979).
\bibitem{Yang1993} B. Yang, K. J. Schafer, B. Walker, K. C. Kulander, P. Agostini, and L. F. DiMauro, Intensity-dependent scattering rings in high order above-threshold ionization, Phys. Rev. Lett. \textbf{71}, 3770 (1993).
\bibitem{Paulus1994} G. G. Paulus, W. Becker, W. Nicklich, and H. Walther, Rescattering effects in above-threshold ionization: a classical model, J. Phys. B \textbf{27}, L703 (1994).
\bibitem{Lewenstein1995} M. Lewenstein, K. C. Kulander, K. J. Schafer, and P. H. Bucksbaum, Rings in above-threshold ionization: A quasiclassical analysis, Phys. Rev. A \textbf{51}, 1495 (1995).
\bibitem{Becker2002} W. Becker, F. Grasbon, R. Kopold, D. B. Milo\u{s}evi\'{c}, G. G. Paulus, and H. Walther,
Above-threshold ionization: from classical features to quantum effects, Adv. At. Mol. Opt. Phys. \textbf{48}, 35 (2002).

\bibitem{McPherson1987} A. McPherson, G. Gibson, H. Jara, U. Johann, T. S. Luk, I. A. McIntyre, K. Boyer, and C. K. Rhodes,
Studies of multiphoton production of vacuum-ultraviolet radiation in the rare gases, J. Opt. Soc. Am. B \textbf{4}, 595(1987).
\bibitem{Huillier1991} A. L'Huillier, K. J. Schafer, and K. C. Kulander, Theoretical aspects of intense field harmonic generation, J. Phys. B \textbf{24}, 3315 (1991).
\bibitem{Corkum} P. B. Corkum, Plasma perspective on strong field multiphoton ionization, Phys. Rev. Lett. \textbf{71}, 1994 (1993).
\bibitem{Lewenstein1994} M. Lewenstein, Ph. Balcou, M. Yu. Ivanov, A. L'Huillier, and P. B. Corkum, Theory of high-harmonic generation by low-frequency laser fields, Phys. Rev. A \textbf{49}, 2117 (1994).


\bibitem{Krausz} P. B. Corkum and F. Krausz, Attosecond science, Nature Phys. \textbf{3}, 381 (2007).
\bibitem{Krausz2009} F. Krausz and M. Ivanov, Attosecond physics, Rev. Mod. Phys. \textbf{81}, 163 (2009).
\bibitem{Vrakking} F. L\'{e}pine, M. Y. Ivanov, and M. J. J. Vrakking, Attosecond molecular dynamics: fact or fiction? Nature Photon. \textbf{8}, 195 (2014).




\bibitem{Xie} X. J. Xie, C. Chen, G. G. Xin, J. Liu, and Y. J. Chen, Coulomb-induced ionization time lag after electrons tunnel out of a
barrier, Opt. Express \textbf{28}, 33228 (2020).

\bibitem{Milosevic2006} D. B. Milo\v{s}evi\'{c}, G. G. Paulus, D. Bauer and W. Becker, Above-threshold ionization by few-cycle pulses, J. Phys. B \textbf{39}, R203 (2006).
\bibitem{Faisal} F. H. M. Faisal, Multiple absorption of laser photons by atoms, J. Phys. B \textbf{6}, L89 (1973).
\bibitem{Reiss} H. R. Reiss, Effect of an intense electromagnetic field on a weakly bound system, Phys. Rev. A \textbf{22}, 1786 (1980).





\bibitem{MishaY} T. Brabec, M. Yu. Ivanov, and P. B. Corkum, Coulomb focusing in intense field atomic processes, Phys. Rev. A \textbf{54}, R2551 (1996).
\bibitem{Goreslavski} S. P. Goreslavski, G. G. Paulus, S. V. Popruzhenko, and N. I. Shvetsov-Shilovski, Coulomb Asymmetry in Above-Threshold Ionization, Phys. Rev. Lett. \textbf{93}, 233002 (2004).
\bibitem{yantm2010} T. M. Yan, S. V. Popruzhenko, M. J. J. Vrakking, and D. Bauer, Phys. Rev. Lett. \textbf{105}, 253002 (2010).









\bibitem{Wang2020} S. Wang, J. Y. Che, C. Chen, G. G. Xin, and Y. J. Chen, Tracing the origins of an asymmetric momentum distribution for polar molecules
in strong linearly polarized laser fields, Phys. Rev. A \textbf{102}, 053103 (2020).

\bibitem{Che2021} J. Y. Che, C. Chen, S. Wang, G. G. Xin, and Y. J. Chen, 
Characterizing Sub-Cycle Electron Dynamics of Polar Molecules by Asymmetry in Photoelectron Momentum Distributions, arXiv:2103.13703 (2021).


\bibitem{Wustelt} P. Wustelt, F. Oppermann, L. Yue, M. M\"{o}ller, T. St\"{o}hlker,
M. Lein, S. Gr\"{a}fe, G. G. Paulus, and A. M. Sayler, Heteronuclear Limit of Strong-Field Ionization: Fragmentation
of HeH$^+$ by Intense Ultrashort Laser Pulses, Phys. Rev. Lett. \textbf{121}, 073203 (2018).

\bibitem{Liwy} W. Y. Li, S. J. Yu, S. Wang, and Y. J. Chen, Probing nuclear dynamics of oriented HeH$^+$ with odd-even high harmonics, Phys. Rev. A \textbf{94}, 053407 (2016).


\bibitem{Feit} M. D. Feit, J. A. Fleck, Jr., and A. Steiger, Solution of the Schr\"{o}dinger Equation by a Spectral Method, J. Comput. Phys. \textbf{47}, 412 (1982).
\bibitem{Wang2017} S. Wang, J. Cai, and Y. J. Chen, Ionization dynamics of polar molecules in strong elliptical laser fields, Phys. Rev. A \textbf{96}, 043413 (2017).



\bibitem{Dimitrovski} D. Dimitrovski, C. P. J. Martiny, and L. B. Madsen, Strong-field ionization of polar molecules: Stark-shift-corrected strong-field approximation, Phys. Rev. A. \textbf{82}, 053404 (2010).






\bibitem{Eckle} P. Eckle, A. N. Pfeiffer, C. Cirelli, A. Staudte, R. D?rner, H. G. Muller, M. B\"{u}ttiker, and U. Keller, Attosecond
ionization and tunneling delay time measurements in helium, Science \textbf{322}, 1525 (2008).




\bibitem [\dag] {1} The authores contribute samely to this paper.
\bibitem [*] {2} chenyjhb@gmail.com


\end{thebibliography}
\end{document}